\documentstyle[psfig,sprocl]{article}
\def\be{\begin{equation}}
\def\ee{\end{equation}}
\def\bea{\begin{eqnarray}}
\def\eea{\end{eqnarray}}

\newcommand{\Be}{\begin{equation}}
\newcommand{\Ee}{\end{equation}}
\newcommand{\Bea}{\begin{eqnarray}}
\newcommand{\Eea}{\end{eqnarray}}
\newcommand{\n}{\nu}
\newcommand{\nbar}{\bar\nu}

\begin{document}
\title{MSW WITHOUT MATTER}
\author{ T. GOLDMAN }
\address{Los Alamos National Laboratory\\Los Alamos, NM 87545, USA}
\author{ B. H. J. MCKELLAR }
\address{University of Melbourne\\Parkville, Victoria 3052, AUSTRALIA}
\author{G. J. STEPHENSON JR.}
\address{University of New Mexico\\ Albuquerque, NM 87131, USA}

\maketitle\abstracts{ We examine the effects of a scalar field,
coupled only to neutrinos, on oscillations
among weak interaction current eigenstates.
The existence of a real scalar field is
manifested as effective masses for the
neutrino mass eigenstates, the same for
$\nbar$ as for $\n$.  Under some conditions,
this can lead to a vanishing of $\delta m^2$,
giving rise to MSW-like effects.  We present
an idealized example and show that it may be possible
to resolve the apparent discrepancy in spectra
required by r-process nucleosynthesis in the
mantles of supernovae and by Solar neutrino
solutions.}

We have recently examined~\cite{Clouds}
the possibility that, in addition to the 
Standard Model interactions, neutrinos interact 
with each
other through an extremely light 
scalar
field $\phi$.
The neutrinos with mass $m_i$ couple to
$\phi$ with constants $g_i$.  We showed 
that, consistent
with known phenomena,
neutrino clouds could form in the 
early
Universe, influence the
evolution of structures on stellar scales, and 
have observable consequences.  Here, we discuss
a consequence of this scalar interaction
that can occur whether clouds form or not.
 
Following the relativistic many-body theory
known as
Quantum Hadrodynamics (QHD)~\cite{QHD},
we define effective masses
\begin{equation}
m^*_j = m_j -g_j\phi
\end{equation}
With more than one mass eigenstate, it is possible 
for some $m_j^*$ to become negative.
The richness of the system can be demonstrated 
with a spherically symmetric model
in which the
various couplings are all equal to the same
constant $g$.  Consider, for simplicity, two 
mass eigenstates, let the vacuum mass of the 
heavier be
denoted by $m_h$ and that of the lighter by
$m_l$.
In this case, the
shift from the vacuum mass to the effective 
mass is the same for both neutrinos,
\Bea
\Delta m &=& g\phi \\
m^*_h &=& m_h - \Delta m\\
m^*_l &=& m_l - \Delta m
\Eea
For large enough shift
 this can lead to $m_l$
becoming very negative.  If
\Bea
m_l^* & = & -m_h^*,\quad\mbox{then}\\
{m_h^*}^2-{m_l^*}^2 & = & 0,
\Eea
and there is a degeneracy between the two
neutrinos arising from a very 
different mechanism
than that involved in the usual MSW
effect~\cite{msw}.  Since the
change in the effective mass is due
to a scalar interaction, it is the same
for both $\n$ and $\nbar$ and the
degeneracy will occur at same
density, hence the same
radius in a star, supernova or
other object, for both.

In the presence of  matter,
there is also a normal
MSW effect which, being an energy
shift due to a vector interaction,
has the opposite sign for $\n$ and
$\nbar$, hence degeneracies
will occur
at different radii.

To illustrate these points we have
generated the cartoons in Figure 1
\begin{figure}[h]
\begin{minipage}{2.3in}
\psfig{file=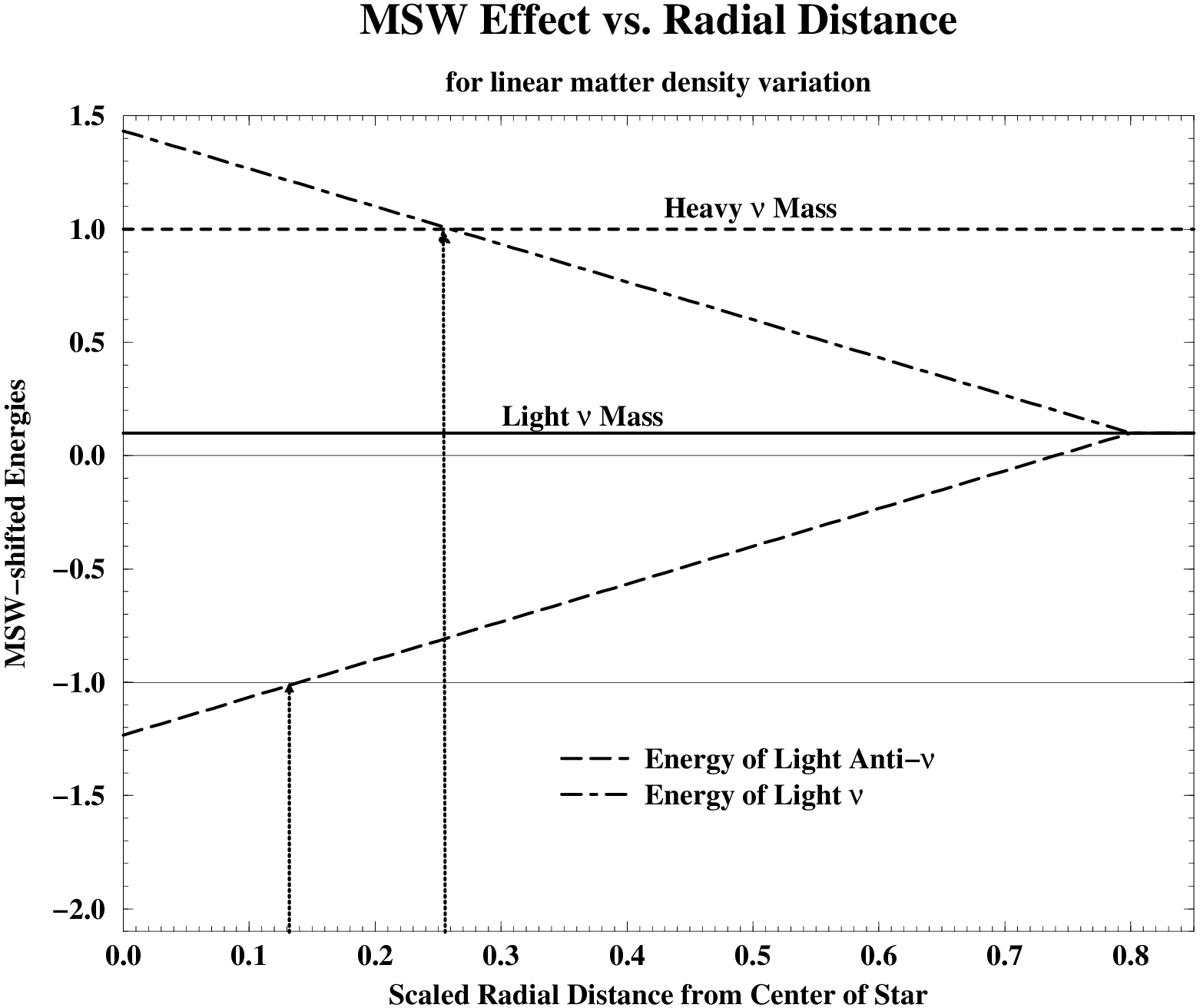,width=2.2in,height=2.5in}
\end{minipage} \hfill
\begin{minipage}{2.3in}
\psfig{file=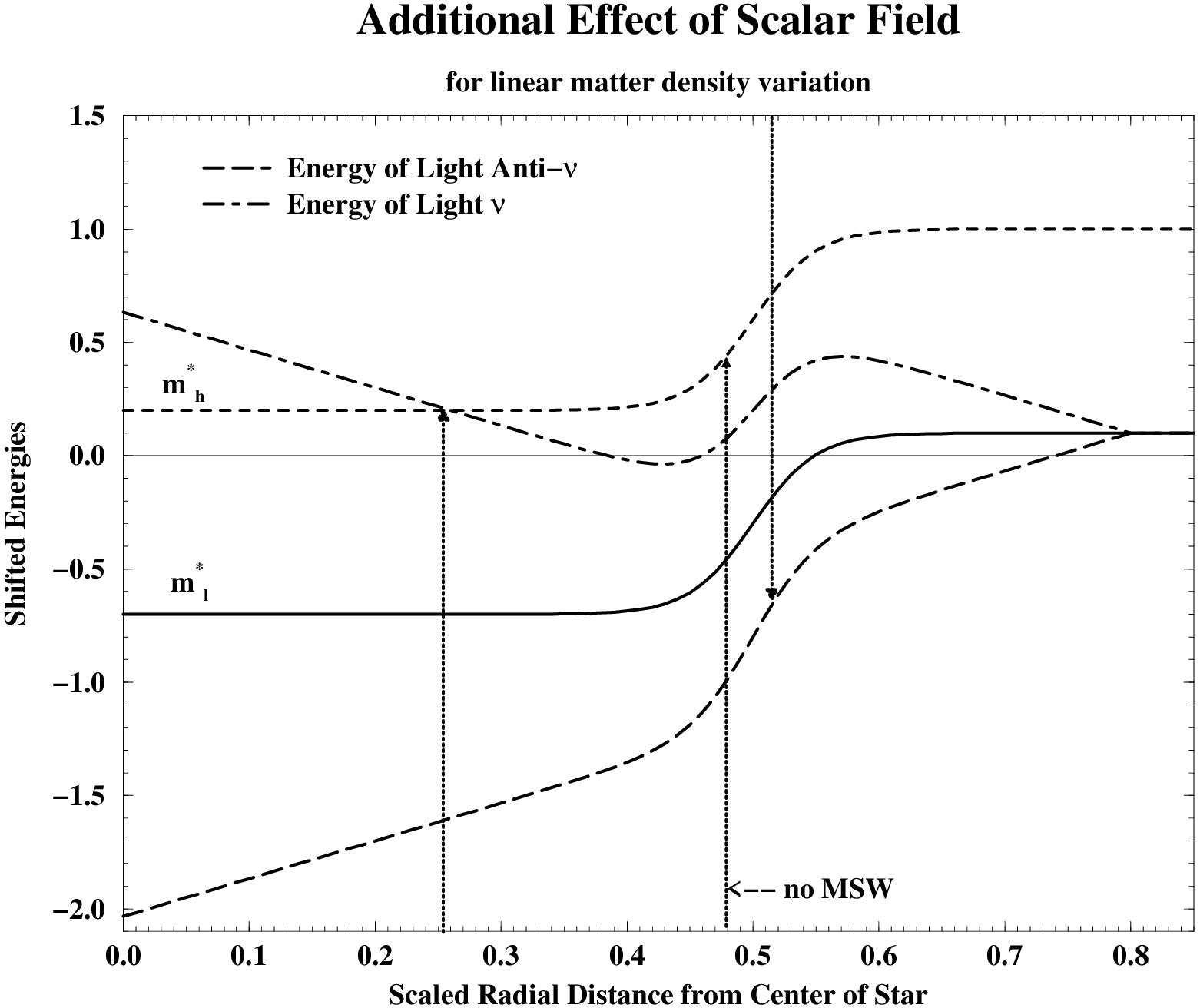,width=2.2in,height=2.5in}
\end{minipage}
\caption{Cartoon of effective mass shifts
and energy shifts described in the text}
\end{figure}
by representing the results of solving
the nonlinear differential equation for
the selfconsistent effective
mass~\cite{Clouds}
with a simple analytic form
and assuming a linear effect
for the vector MSW (clearly, this is far
too simple for a real system, but the trends
are correctly represented).  In part A we
assume no scalar field and demonstrate 
that the $\nbar$ degeneracy, indicated by
the shorter vertical line, occurs at
a smaller radius than
the $\n$ degeneracy, indicated by the 
longer vertical line.   In B, we add the
scalar.  The middle vertical line indicates
the position of the ${m^*}^2$ degeneracy
ignoring the
vector MSW; the outer vertical line
indicates the position of the $\nbar$
degeneracy with both.  The position of
the $\n$ degeneracy is shown by the 
inner most vertical line.

This result has possible physical 
implications.  It has recently been
shown~\cite{Fuller} that r-process
nucleosynthesis in the exterior of
a supernova can give a credible 
account for abundances, provided
there is an excess of neutrons over
protons.  To achieve this, it is
desirable to have the $\nbar$ at a 
higher temperature than the $\n$
at the site of the r-process,
which can be achieved through 
enhanced flavor transitions if the
$\nbar$ transition occurs outside
the $\n$ transition~\cite{Fuller}.
These authors suggest that this
can be achieved by an inverted 
spectrum ($m_{\n_e}$ larger 
than some other mass); it could
also be achieved through a scalar
interaction.

The extension of these considerations
to three generations is straightforward
and will be presented 
elsewhere~\cite{mswwom}.

This work has been supported in part by the
United States National Science Foundation, 
the United States Department of Energy,
the Australian Research Council and the
Australian DIST.

\end{document}